\documentclass{article}
\usepackage{graphicx} 
\usepackage{setspace}
\usepackage{dcolumn} 
\usepackage[T1]{fontenc}
\usepackage[sc]{mathpazo}

\usepackage{bm} 
\usepackage{graphicx}
\usepackage{amsmath}  
\usepackage{amsthm}   
\usepackage{enumitem}
\usepackage{latexsym}
\usepackage{amsfonts}   
\usepackage{amssymb}
\usepackage{array}      
\usepackage{epsfig}
\usepackage{braket} 
\usepackage{bbold}
\usepackage{color}
\usepackage[colorlinks=true,linkcolor=blue,urlcolor=blue,citecolor=blue,pdfusetitle]{hyperref}
\usepackage{authblk} 
\usepackage{cancel}
\usepackage{tikz}
\usepackage{algorithm}
\usepackage{algpseudocode}
\newcommand{\QED}{\hfill\ensuremath{\square}}
\newcommand{\eq}{Eq.~}

\DeclareRobustCommand\openzero{\leavevmode\hbox{0\kern-.55em0}}

\newtheorem{definition}{Definition}
\newtheorem{remark}{Remark}
\newtheorem{lemma}{Lemma}

\title{Commentary on the decomposition of universal multiport interferometers: how it works in practice}
\author[1]{Dario Cilluffo
\thanks{dario.cilluffo@uni-ulm.de}}
\affil{Institut f\"ur Theoretische Physik und IQST, Albert-Einstein-Allee 11, Universit\"at Ulm, D-89081 Ulm, Germany
}

\date{December 2024}

\begin{document}

\maketitle

\begin{abstract}
The decomposition of multiport interferometers is a fundamental tool in quantum optics and computing. This note aims to serve as a concise reference for performing the decomposition according to the most common design approaches, offering a self-contained treatment with essential mathematical details and practical working examples. Specifically, we provide a clear, self-consistent presentation of all the mathematics required to determine the gate parameters and apply the necessary phase corrections to achieve the desired structure.
\end{abstract}

\section{Introduction}
Any noiseless multiport interferometer can be expressed as a structured configuration of beam splitters and phase shifters. Among the most significant decomposition schemes are the Reck design \cite{reck1994experimental} and the Clements design \cite{clements2016optimal}. These approaches reconstruct the interferometer's unitary scattering matrix, $U$, through a systematic sequence of transformations involving special unitary matrices that act on pairs of modes. Each transformation is crafted to eliminate a specific target element in $U$ by mixing corresponding rows or columns. This iterative process ultimately yields a diagonal matrix consisting solely of phase factors. By reversing the sequence, one can decompose $U$ into a product of these special unitary operations.
Despite their broad adoption in quantum optics and interferometry, practical implementation of these designs, as described in \cite{reck1994experimental} and \cite{clements2016optimal}, is often non-trivial. Critical aspects, such as the precise selection of two-mode unitaries, are often left implicit.

In this notes, we provide a comprehensive, step-by-step guide to constructing a Clements decomposition algorithm for an arbitrary interferometer matrix. We begin by establishing a solid mathematical foundation for the method and proceed with detailed instructions on constructing the necessary nullifying left and right matrices. Furthermore, we discuss the final adjustments for phase factors and propose an efficient approach for storing the decomposition data.
To facilitate implementation, we include pseudo-code summarizing the entire decomposition procedure. In the main text, we detail the steps for decomposing a unitary matrix following the Clements design, ensuring clarity by working directly with numerical examples rounded to the tenth significant digit. This precise treatment aims to serve as a practical reference for students and researchers undertaking such a decomposition for the first time.

\section{The mathematical framework: Givens rotations}
Givens rotations, named after Wallace Givens \cite{givens1958computation}, were introduced in the 50's for simplifying numerical problems involving matrices like triangulation and QR decomposition, and are now part of the standard toolkit of quantum circuit theory \cite{cybenko2001reducing}. We provide here an operational definition of Givens rotation matrices:

\begin{definition}{Givens rotation.}
\label{def:G}
Let $\mathbf{x} \in \mathbb{C}^2$ with $r =\Vert\mathbf{x}\Vert$. Then a Givens rotation is a matrix $G\in SU(2)$ such that  $ G\cdot \mathbf{x} = \left(\begin{matrix} \tilde{r} \\0 \end{matrix}\right)$,
with $|\tilde{r} | = r$.
\end{definition}
In the real case, Givens rotation reduce to counterclockwise rotations of the vector $\mathbf{x}$. Notice that $G$ is not unique and there are multiple ways to parametrize it. 
The outcome of the operation is the elimination of one element (the lower in our case) of the vector $\mathbf{x}$.

\begin{remark}
If instead of $\mathbf{x}$ we consider a 2-row matrix $A$, the matrix $G$ in {\rm Def.\ref{def:G}} can be used to eliminate any element in the second row of $A$.
\end{remark}
In other words, $G$ can be understood as an operation that mixes the rows of a given $2\times M$ matrix in order to nullify a target element in its second row.
When $G$ is embedded into a $M\times M$ identity matrix, it can be applied to a $M$-dimensional vector or an $M$-row matrix in the same way. We denote an embedded Givens rotation by $\mathcal{G}$.
Thus, specifically, embedded Givens rotations can be used to triangulate a generic complex matrix.

The attentive reader may have already guessed the mechanism empowering Givens rotations to decompose a unitary via triangulation:

\begin{lemma}
A triangular unitary matrix $A$ is a diagonal matrix with $|A_{ii}|=1$.
\end{lemma}
\textit{Proof.} Assume $A$ is lower-triangular. Thus $A\cdot A^\dag = I \Rightarrow |A_{00}|^2=1$ and $A_{0j} \cdot A_{0j}^* =0~~\forall j\neq 0 \Rightarrow A_{j 0} = 0 ~~ \forall j\neq 0 $. Iterating for all the rows gives the result. If $A$ is upper triangular the same procedure applies by considering $A^\dag\cdot A=I$. From the unitarity it follows $|A_{ii}|=1$.
\QED

Thus we can use a product of Givens rotations to upper or lower triangulate $\mathcal{U}$, having as a result a diagonal matrix containing only phase factors. Therefore by inverting the triangulation protocol we have back a decomposition of $\mathcal{U}$ as a collection of $SU(2)$ matrices (beam splitters) plus a phase shifts.
This decomposition is not unique, and the choice of the rotation pattern plays a critical role. Mixing two rows after eliminating certain elements can potentially reintroduce nonzero values in the previously zeroed entries.
In the next section we will deepen the definition of the rotation and introduce left and right rotations to overcome this problem.

\subsection{Left and right rotations}

The principles discussed in the previous section provide sufficient insight into how the Reck and Clements design operates. However, we will now delve into the nuances of applying Givens rotations in a typical scenario. To do so, we will refine and slightly modify the earlier definitions to align with the specific forms of beam splitters and phase shifters we intend to use. Consequently, the following discussion is heavily influenced by the form of the local transformations we choose to describe the optical elements.

We define the left givens rotation with the form of a beam splitter matrix as defined in \cite{haroche2006exploring}.
\begin{align}
G_{\theta, \varphi} := \left( \begin{matrix}
\cos(\theta/2) & i e^{-i \varphi}\sin(\theta/2) \\
i e^{i \varphi} \sin(\theta/2) & \cos(\theta/2)
\end{matrix}
\right) \equiv \left( \begin{matrix}
a & i b \\
i b^* & a
\end{matrix} 
\right) =: G_{a,b}
\end{align}
According to Def.\ref{def:G}, we find the values of $a$ and $b$ that eliminate the lower element of a bi-vector $\mathbf{x}$: 

\begin{align}
\left( \begin{matrix}
a & i b \\
i b^* & a
\end{matrix}
\right) \cdot
\left( \begin{matrix}
x_1 \\
x_2
\end{matrix}
\right) = \left( \begin{matrix}
r \frac{x_1}{|x_1|} \\ 0
\end{matrix}
\right) \Rightarrow
a= \frac{|x_1|}{r},  ~~b= - i \frac{x_2^*}{r} \frac{x_1}{|x_1|}\,,
\end{align}
where we replaced $\tilde{r} =  r \frac{x_1}{|x_1|}$ in order to keep $a$ real.
For the sake of clarity we present here an example of application on a symmetric tritter (the normalized discrete Fourier transform of dimension $3$):
\begin{align}
U =
\left(
\everymath{\scriptstyle}
\begin{matrix}
 0.57735027 &  0.57735027 &  0.57735027 \\
  0.57735027 & -0.28867513 - 0.5i & -0.28867513 + 0.5i \\
  0.57735027  & -0.28867513 + 0.5i & -0.28867513 - 0.5i
\end{matrix}
\right)
\end{align}
We want to eliminate the bottom-left element, $U_{2,0}$, thus $x_1=U_{1,0}$ and $x_2=U_{2,0}$. The rotation $\mathcal{G} = 1 \oplus G_{a,b}$, with $a=1/\sqrt{2}$ and $b=-i/\sqrt{2}$ mixes the last two columns to achieve the goal:
\begin{align}
\mathcal{G}\cdot U =
\left(
\everymath{\scriptstyle}
\begin{matrix}
 0.57735027 &  0.57735027 &  0.57735027 \\
  0.81649658  & -0.40824829 & -0.40824829 \\
  0  & 0.70710678i & -0.70710678i
\end{matrix}
\right) = U\,,
\end{align}
where, with an abuse of notation, we redefined $U$ as the result of the rotation.
As expected, the elements of the first line remain unchanged. Note that $\mathcal{G}$ is a gate acting on the modes $1$ and $2$ of our interferometer. 
The route to the low-triangular matrix requires now to eliminate $U_{1,0}$ or $U_{2,1}$. The elimination of $U_{2,1}$ would require mixing again the lines $1$ and $2$ but this can in general create a nonzero value in $U_{2,0}$, unless $U_{1,0}=0$. Thus we can proceed by nullifying $U_{1,0}$ first, with a rotation $\mathcal{G} = G_{a,b} \oplus 1$ and finally $U_{2,0}$.
This sequential strategy aligns with the Reck design for the target interferometer, as detailed in \cite{reck1994experimental}.
Rather than completing the calculation, we shift our focus to the Clements scheme, which makes use of rotations applied from both the left and the right. A matrix element can be eliminated through right multiplication by a rotation matrix, with the key distinction being that this operation mixes two columns of the matrix.
If we try to nullify $U_{2,0}$ via right multiplication we immediately notice that we must apply a gate on the modes $0$ and $1$, i.e.~$\mathcal{G}^\dag=G_{a,b} \oplus 1$, as we this mixes the first two columns.  For convenience, matrices multiplied from the right are defined as the adjoint of $\mathcal{G}$. Alternatively, this can be understood by observing that right multiplication corresponds to left multiplication on the adjoint of $U$, such that $U\cdot \mathcal{G}^\dag = (\mathcal{G} \cdot U^\dag )^\dag$. As a consequence, if we aim to eliminate $U_{2,0}$, we need to act on the first two lines of $U^\dag$, and the element to be nullified will be $U^\dag_{0,2}$ and the second element of the bi-vector can only be $U^\dag_{1,2}$.
Keeping this in mind, we define right Givens rotations as 
\begin{definition}{Right Givens rotation.}
\label{def:Gr}
Let $\mathbf{x} \in \mathbb{C}^2$ with $\Vert\mathbf{x}\Vert=r$. Then a left Givens rotation is a matrix $G\in SU(2)$ such that  $ G\cdot \mathbf{x} = \left(\begin{matrix} 0 \\ \tilde{r} \end{matrix}\right)$, with $|\tilde{r}| = r$
\end{definition}
Thus we can now define our (embedded) right rotation $\mathcal{G}^\dag=G_{a,b}\oplus 1$, $a=1/\sqrt{2}$, $b=-i/\sqrt{2}~e^{-i 2 \pi/3 }$ and eliminate $U_{2,0}$ via right multiplication with $\mathcal{G}^\dag$:
\begin{align}
 U \cdot \mathcal{G}^\dag =
\left(
\everymath{\scriptstyle}
\begin{matrix}
 0.61237244+0.35355339i &  0.20412415+0.35355339i &  0.57735027 \\
  0.61237244-0.35355339i  & -0.40824829 & -0.28867513+0.5i \\
  0  & -0.40824829+0.70710678i & -0.28867513-0.5i
\end{matrix}
\right) = U\,,
\end{align}
As expected, the third column remains unchanged. We can now decide how to proceed to upper triangulate $U$: again, we could continue with right multiplications or with left multiplications. As there's no difference between the Reck and Clements designs for a 3-mode interferometer, in the next section we will follow step by step the decomposition of a symmetric 4-mode interferometer (\textit{quatter}) according to the Clements design.

\section{Decomposition of a symmetric quatter}

The target matrix is, for the sake of clarity, the symmetric quatter (normalized discrete Fourier transform of dimension $4$):
\begin{align}
U =
\left(
\everymath{\scriptstyle}
\begin{matrix}
 0.5 &  0.5 &  0.5 &  0.5 \\
  0.5 & - 0.5i & -0.5 & 0.5i \\
  0.5  &  -0.5 & 0.5  & -0.5 \\
  0.5  &  0.5i & - 0.5 & -0.5i
\end{matrix}
\right)
\end{align}
We choose to upper triangulate $U$ by starting from the elimination of $U_{3,0}$ by right multiplication, i.e.~by mixing two columns, which must be necessarily the first two ones.
Thus $\mathcal{G}_{0,1}^\dag=G_{a,b}\oplus I_2$, where the subscript to $\mathcal{G}$ indicates the modes we are acting on and $I_2$ is the 2-dimensional identity matrix.
In this case $a=1/\sqrt{2},~b=-1/\sqrt{2} $, and we obtain

\begin{align}
U\cdot \mathcal{G}_{0,1}^\dag =
\left(
\everymath{\scriptstyle}
\begin{matrix}
 0.35355339+0.35355339i &  0.35355339+0.35355339i &  0.5 &  0.5 \\
  0.70710678 & 0 & -0.5 & 0.5i \\
  0.35355339-0.35355339i & -0.35355339+0.35355339i & 0.5  & -0.5 \\
  0 &  0.70710678 & - 0.5 & -0.5i
\end{matrix}
\right)
\end{align}
We now proceed nullifying $U_{2,0}$ through left multiplication: the embedded rotation will be $\mathcal{G}_{1,2} = 1\oplus G_{a,b} \oplus 1$, $a=\sqrt{2/3}, b=\sqrt{1/3}e^{-i \pi/4}$ and
\begin{align}
&\mathcal{G}_{1,2} \cdot U\cdot \mathcal{G}_{0,1}^\dag \notag\\&=
\left(
\everymath{\scriptstyle}
\begin{matrix}
 0.35355339+0.35355339i &  0.35355339+0.35355339i &  0.5 &  0.5 \\
  0.8660254 & -0.28867513 & -0.20412415+0.20412415i & -0.20412415+0.20412415 \\
  0 & -0.28867513+0.28867513i & 0.61237244-0.20412415i  & -0.61237244-0.20412415i \\
  0 &  0.70710678 & - 0.5 & -0.5i
\end{matrix}
\right)\,.
\end{align}
We now aim to eliminate the element $U_{3,1}$. The most natural choice is to proceed with left multiplication as the left elements of the last two lines are already $0$.
Thus we apply the rotation
$\mathcal{G}_{2,3} = I_2\oplus G_{a,b}$, $a=1/2, b=\sqrt{3/4}e^{-i \pi/4}$:
\begin{align}
&\mathcal{G}_{2,3}\cdot\mathcal{G}_{1,2} \cdot U\cdot \mathcal{G}_{0,1}^\dag \notag\\&=
\left(
\everymath{\scriptstyle}
\begin{matrix}
 0.35355339+0.35355339i &  0.35355339+0.35355339i &  0.5 &  0.5 \\
  0.8660254 & -0.28867513 & -0.20412415+0.20412415i & -0.20412415+0.20412415 \\
  0 & -0.57735027+0.57735027i & -0.40824829i  & -0.40824829i \\
  0 &  0 & - 0.5 + 0.5i & 0.5-0.5i
\end{matrix}
\right)\,.
\end{align}
Once again, there is no mandatory choice for how to proceed; however, the guiding principle is to avoid altering elements that have already been zeroed. With this in mind, a sensible approach is to mix the last two columns to eliminate $U_{3,2}$. Thus, 
we have 
$\mathcal{G}_{2,3}^\dag = I_2\oplus G_{a,b}$, $a=\sqrt{1/2}, b=i \sqrt{1/2}$:
\begin{align}
&\mathcal{G}_{2,3}\cdot\mathcal{G}_{1,2} \cdot U\cdot \mathcal{G}_{0,1}^\dag\cdot \mathcal{G}_{2,3}^\dag \notag\\&=
\left(
\everymath{\scriptstyle}
\begin{matrix}
 0.35355339+0.35355339i &  0.35355339+0.35355339i &  0.70710678 &  0 \\
  0.8660254 & -0.28867513 & -0.28867513+0.28867513i & 0 \\
  0 & -0.57735027+0.57735027i & -0.57735027i  & 0 \\
  0 &  0 & 0 & 0.70710678-0.70710678i
\end{matrix}
\right)\,.
\end{align}
Now we are in the same situation as before but transposed: the two bottom elements of the middle columns are zero, thus it is convenient to mix those columns to eliminate the $U_{2,1}$. This requires a further right multiplication 
by $\mathcal{G}_{1,2}^\dag = 1\oplus G_{a,b} \oplus 1$, $a=\sqrt{1/3}, b=-\sqrt{2/3} e^{-\pi/4}$:
\begin{align}
&\mathcal{G}_{2,3}\cdot\mathcal{G}_{1,2} \cdot U\cdot \mathcal{G}_{0,1}^\dag\cdot \mathcal{G}_{2,3}^\dag \cdot\mathcal{G}_{1,2}^\dag \notag\\&=
\left(
\everymath{\scriptstyle}
\begin{matrix}
 0.35355339+0.35355339i &  0.61237244+0.61237244i &  0 &  0 \\
  0.8660254 & 0.5  & 0 & 0 \\
  0 & 0 & -i  & 0 \\
  0 &  0 & 0 & 0.70710678-0.70710678i
\end{matrix}
\right)\,.
\end{align}
We conclude by eliminating $U_{1,0}$ with the last right-multiplication by
$\mathcal{G}_{0,1}^\dag = G_{a,b} \oplus I_2$, $a=1/2, b=i \sqrt{3/4}$, which gives
\begin{align}
&\mathcal{G}_{2,3}\cdot\mathcal{G}_{1,2} \cdot U\cdot \mathcal{G}_{0,1}^\dag\cdot \mathcal{G}_{2,3}^\dag \cdot\mathcal{G}_{1,2}^\dag \cdot\mathcal{G}_{0,1}^\dag \notag\\&=
\left(
\everymath{\scriptstyle}
\begin{matrix}
 e^{i\pi/4} &  0 &  0 &  0 \\
  0 & -1  & 0 & 0 \\
  0 & 0 & -i  & 0 \\
  0 &  0 & 0 & e^{-i\pi/4}
\end{matrix} 
\right) = U_\phi\,,
\end{align}
which is a diagonal matrix containing only phase factors we denote as $U_\phi$.
This procedure already gives gate-wise decomposition of our interferometer:
\begin{align}
 U = \mathcal{G}_{1,2}^\dag\cdot\mathcal{G}_{2,3}^\dag \cdot U_\phi \cdot\mathcal{G}_{0,1} \cdot\mathcal{G}_{1,2} \cdot \mathcal{G}_{2,3} \cdot\mathcal{G}_{0,1}
 \label{eq:decomp}
\end{align}
The beam-splitter angles and phases are retrieved from all the couples $(a,b)$ we collected. The parameters we found are reported in Table \ref{tab:4mode1}.

\begin{table}[h!]
\centering
\begin{tabular}{|c|c|c|}
\hline
\textbf{Modes} & \textbf{$a$} & \textbf{$b$} \\
\hline
$(1, 2)$ & $1/\sqrt{2}$ & $-1/\sqrt{2}$ \\
$(2, 3)$ & $\sqrt{2/3}$ & $\sqrt{1/3}e^{-i \pi/4}$ \\
$(0, 1)$ & $1/2$ & $\sqrt{3/4}e^{-i \pi/4}$ \\
$(1, 2)$ & $\sqrt{1/2}$ & $i \sqrt{1/2}$ \\
$(2, 3)$ & $\sqrt{1/3}$ & $-\sqrt{2/3} e^{-\pi/4}$ \\
$(0, 1)$ & $1/2$ & $i \sqrt{3/4}$ \\
\hline
\end{tabular}
\caption{Values of $a=G_{00}$, and $b=G_{01}$ for the gates in \eqref{eq:decomp} -from left to right-.}
\label{tab:4mode1}
\end{table}
Before proceeding to the final refinements, we note that the elimination pattern we followed was not arbitrary. Starting from the bottom-left corner, we progressed to the nearest subdiagonal, moving from top to bottom, then reversed direction to traverse the next subdiagonal from bottom to top, continuing this zigzag—or \textit{boustrophedon}\footnote{from the Ancient Greek "like the ox turns [while plowing]."}—path until reaching the final subdiagonal. This pattern is directly reflected in the placement of the gates within the interferometer (see Fig.~\ref{fig:fig1} for an illustrative example).

\subsection{Refinements and sorting}
Our matrix is now decomposed as a train of rotations and local phase shifts. We want now to put the phases on the left of the rotations. 
Thus we turn \eq\eqref{eq:decomp} into
\begin{align}
 U = U_\phi \cdot (U_\phi^\dag \cdot \mathcal{G}_{1,2}^\dag \cdot U_\phi ) \cdot (U_\phi^\dag \cdot \mathcal{G}_{2,3}^\dag \cdot U_\phi )\cdot\mathcal{G}_{0,1} \cdot\mathcal{G}_{1,2} \cdot \mathcal{G}_{2,3} \cdot\mathcal{G}_{0,1}\,.
 \label{eq:decomp2}
\end{align}
If we denote with $\phi_1$ and $\phi_2$ the phases of the diagonal elements of $U_\phi$ corresponding to the non-identity block in each $\mathcal{G}$ we note that 
\begin{align}
\left( \begin{matrix}
e^{-i \phi_1} & 0 \\
0 & e^{-i \phi_2}
\end{matrix}
\right) \cdot \left( \begin{matrix}
a & -i b^* \\
-i b & a
\end{matrix}
\right) \cdot 
\left( \begin{matrix}
e^{i \phi_1} & 0 \\
0 & e^{i \phi_2}
\end{matrix}
\right) = \left( \begin{matrix}
a & -i b^* e^{-i (\phi_1-\phi_2)} \\
-i b e^{i (\phi_1-\phi_2)} & a
\end{matrix}
\right)\,
\end{align}
which shows that we can simply make the following replacement 
\begin{align}
\left( \begin{matrix}
a & -i b^* \\
-i b & a
\end{matrix}
\right) \rightarrow \left( \begin{matrix}
a & i (b^* e^{-i (\phi_1-\phi_2 -\pi)}) \\
i (b^* e^{-i (\phi_1-\phi_2 -\pi)})^* & a
\end{matrix}
\right)\,.
\label{eq:corr}
\end{align}
The decomposition taking into accounts these changes reads
\begin{align}
 U = U_\phi \cdot \mathcal{G}_{1,2} \cdot \mathcal{G}_{2,3} \cdot\mathcal{G}_{0,1} \cdot\mathcal{G}_{1,2} \cdot \mathcal{G}_{2,3} \cdot\mathcal{G}_{0,1}\,.
 \label{eq:decomp_final}
\end{align}
where we maintained $\mathcal{G}_{1,2}$ and $\mathcal{G}_{2,3}$ to mean embedded rotations with the same structure as the old ones but with non-identity element modified according to \eq\eqref{eq:corr}.
For the interferometer under consideration, the parameters are summarized in Table \ref{tab:4mode}.
\begin{figure}
\centering
\includegraphics[scale=0.7,angle=0]{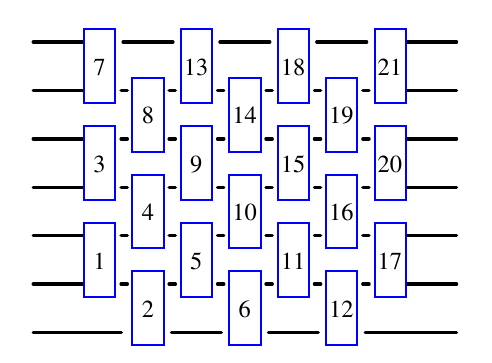}	
\caption{Decomposition of a 7-mode interferometer, with each gate numbered according to its position in the sequence determined by the algorithm.}
\label{fig:fig1}
\end{figure}

\begin{table}[h!]
\centering
\begin{minipage}{0.45\textwidth}
\centering
\begin{tabular}{|c|c|c|}
\hline
\textbf{Modes} & \textbf{$\theta$} & \textbf{$\varphi$}  \\
\hline
$(1, 2)$ & $1.23095942$ & $-2.35619449$ \\
$(2, 3)$ & $2.0943951$ & $3.14159265$ \\
$(0, 1)$ & $2.0943951$ & $-1.57079633$ \\
$(1, 2)$ & $1.91063324$ & $-2.35619449$ \\
$(2, 3)$ & $1.57079633$ & $-1.57079633$ \\
$(0, 1)$ & $1.57079633$ & $ -3.14159265$ \\
\hline
\end{tabular}
\end{minipage}
\begin{minipage}{0.45\textwidth}
\centering
\begin{tabular}{|c|}
\hline
\textbf{$\phi$} \\
\hline
$\pi/4$ \\
$\pi$ \\
$-\pi/2$ \\
$-\pi/4$ \\
\hline
\end{tabular}
\end{minipage}
\caption{Table showing values of $\theta=2 \arccos(G_{0,0})$, and $\varphi = \arg(G_{0,1})$ for each gate in \eqref{eq:decomp_final} -from left to right- and the value of the global phase shifts $\phi$.}
\label{tab:4mode}
\end{table}
The final key aspect to address is how the data from the decomposition is stored. As illustrated in Fig. \ref{fig:fig1}, the decomposition in \eq\eqref{eq:decomp_final} follows the same zig-zag ordering observed earlier, which typically mixes gates from different layers. For instance, the Table \ref{tab:7mode} presents the data from the decomposition of a 7-mode interferometer, associating each gate with its corresponding layer. Thus an extra-sorting might be necessary depending on the applications.

\begin{table}[h!]
\centering
\begin{minipage}{0.4\textwidth}
\centering
\begin{tabular}{|c|c|c|c|}
\hline
\textbf{Layer} & \textbf{Modes} & \textbf{$\theta$} & \textbf{$\varphi$} \\
\hline
1 & $(4, 5)$ & $1.01328373$ & $2.66378246$ \\
2 & $(5, 6)$ & $2.23804657$ & $2.59202681$ \\
1 & $(2, 3)$ & $0.8374462$  & $4.23011761$ \\
2 & $(3, 4)$ & $1.79352577$ & $3.01966744$ \\
3 & $(4, 5)$ & $2.33427509$ & $2.95836058$ \\
4 & $(5, 6)$ & $1.74637704$ & $3.42780842$ \\
1 & $(0, 1)$ & $1.57079633$ & $4.8260573$  \\
2 & $(1, 2)$ & $2.18559956$ & $4.28537678$ \\
3 & $(2, 3)$ & $2.12564842$ & $4.23011761$ \\
4 & $(3, 4)$ & $1.96812101$ & $3.41722926$ \\
5 & $(4, 5)$ & $2.33427509$ & $3.9690119$  \\
6 & $(5, 6)$ & $2.23804657$ & $4.26359003$ \\
3 & $(0, 1)$ & $1.84252123$ & $-2.17184699$ \\
4 & $(1, 2)$ & $2.14816964$ & $-2.68190033$ \\
5 & $(2, 3)$ & $2.12564842$ & $-2.91719318$ \\
6 & $(3, 4)$ & $1.79352577$ & $-2.46839423$ \\
7 & $(4, 5)$ & $1.01328373$ & $-2.01959528$ \\
5 & $(0, 1)$ & $1.84252123$ & $-3.10007209$ \\
6 & $(1, 2)$ & $2.18559956$ & $2.91719318$ \\
7 & $(2, 3)$ & $0.8374462$  & $-2.91719318$ \\
7 & $(0, 1)$ & $1.57079633$ & $2.46839423$ \\
\hline
\end{tabular}
\end{minipage} \hfill
\begin{minipage}{0.38\textwidth}
\centering
\begin{tabular}{|c|}
\hline
\textbf{$\phi$} \\
\hline
$-0.99002554 + 0.14088804i$ \\
$-0.40609043 + 0.9138329i$ \\
$0.4804896 + 0.87700042i$ \\
$0.97053839 - 0.24094653i$ \\
$0.70056012 - 0.71359339i$ \\
$0.13161199 - 0.99130131i$ \\
$-0.42633443 - 0.90456562i$ \\
\hline
\end{tabular}
\end{minipage}
\caption{Layer, modes, $\theta$, and $\varphi$ and global phase shifts $\phi$ for the decomposition of a symmetric 7-mode interferometer.}
\label{tab:7mode}
\end{table}

\clearpage

\section{Algorithm}
We conclude with a possible pseudocode algorithm that implements the concepts discussed in the previous sections.
Clarity and adherence to the descriptions in the main text have been prioritized; however, particularly in the sorting process, more efficient implementations are possible.

\begin{algorithm}[H]
\caption{Clements Decomposition Routine}
\begin{algorithmic}[1]
\Procedure{ClementsDecompose}{$U$}
    \State \textbf{Input}: $U$ (matrix to decompose)
    \State \textbf{Output}: $U_\phi$ (phase shift matrix), $Arguments$ (stored parameters)

    \State \textbf{Global Variables:} $M, Llist, Rlist, Arguments$
    \State $M \gets \text{Number of rows}(U)$
    \State $Llist, Rlist, Arguments, Layers \gets []$ \Comment{Initialize lists}
    
    \State Initialize $i \gets M-1, j \gets 0$ \Comment{Starting indices}
    \State $U \gets \Call{Right Givens}{U, i, j}$
    
    \For{$k \gets 1$ to $M-1$}
        \If{$k$ is odd} \Comment{Odd iterations process the left side}
            \State $i \gets i-1$
            \State $U \gets \Call{Left Givens}{U, i, j}$
            \For{$l \gets 0$ to $k$}
                \State $i \gets i-1$, $j \gets j+1$
                \State $U \gets \Call{Left Givens}{U, i, j}$
            \EndFor
        \Else \Comment{Even iterations process the right side}
            \State $j \gets j+1$
            \State $U \gets \Call{Right Givens}{U, i, j}$
            \For{$l \gets 0$ to $k$}
                \State $i \gets i-1$, $j \gets j-1$
                \State $U \gets \Call{Right Givens}{U, i, j}$
            \EndFor
        \EndIf
    \EndFor
    
    \State $U_\phi \gets U$ \Comment{Final transformed matrix}
    \State $Arguments \gets \Call{Store}{\,}$
    \State \Return $U_\phi, Arguments$
\EndProcedure
\end{algorithmic}
\end{algorithm}

\begin{algorithm}[H]
\caption{Functions: rotations}
\begin{algorithmic}[1]
\Function{Right Givens}{$U,i,j$}
    \State $\mathcal{G} \gets I_M$ \Comment{Initialize identity matrix of size $M$ (number of rows in $U$)}
    \If{$U[i,j]\neq 0$} \Comment{Only proceed if the element is non-zero}
        \State $x_1, x_2 \gets U[i,j]^*,U[i,j+1]^*$
        \State $r \gets \sqrt{|x_1|^2+|x_2|^2}$
        \State $a, b \gets |x_2|/r, i x_2^*/r \cdot x_1/|x_2| $
        \State $\mathcal{G}[j, j] \gets a$         \Comment{Update the rotation matrix $\mathcal{G}$}
        \State $\mathcal{G}[j, j+1] \gets i b$
        \State $\mathcal{G}[j+1, j] \gets i b^*$
        \State $\mathcal{G}[j+1, j+1] \gets a$
        \State \textbf{Append} $[j, 2  \arccos(a), \text{angle}(b)]$ \textbf{to} $Rlist$ \Comment{Collect parameters}
    \EndIf
\State \Return $U \cdot \mathcal{G}^\dag$
\EndFunction

\vspace{1em}

\Function{Left Givens}{$U,i,j$}
    \State $\mathcal{G} \gets I_M$ \Comment{Initialize identity matrix of size $M$ (number of rows in $U$)}
    \If{$U[i,j]\neq 0$} \Comment{Only proceed if the element is non-zero}
        \State $x_1, x_2 \gets U[i-1,j],U[i,j]$
        \State $r \gets \sqrt{|x_1|^2+|x_2|^2}$
        \State $a, b \gets |x_1|/r, -i x_2^*/r \cdot x_1/|x_1| $
        \State $\mathcal{G}[i-1, i-1] \gets a$         \Comment{Update the rotation matrix $\mathcal{G}$}
        \State $\mathcal{G}[i-1, i] \gets i b$
        \State $\mathcal{G}[i, i-1] \gets i b^*$
        \State $\mathcal{G}[i, i] \gets a$
        \State \textbf{Append} $[i-1, 2  \arccos(a), \text{angle}(b)]$ \textbf{to} $Llist$ \Comment{Collect parameters}
    \EndIf
\State \Return $\mathcal{G} \cdot U$
\EndFunction
\end{algorithmic}
\end{algorithm}

\begin{algorithm}[H]
\caption{Functions: Data Sorting and Storing}
\begin{algorithmic}[1]
\Function{Store}{\,}
    \State \textbf{Input}: Global variables $Llist, Rlist, U_\phi$, $M$
    \State \textbf{Output}: $Layers$ (sorted and processed data structure)

    \State Compute $\Delta \phi[l]$ for $l \gets 0$ to $k$: \Comment{Phase angle differences}
    \For{$l \gets 0$ to $k$}
        \State $\Delta \phi[l] \gets \text{angle}(U_\phi[l+1, l+1]) - \text{angle}(U_\phi[l, l])$
    \EndFor

    \State Update phase angles in $Llist$: 
    \ForAll{$entry$ in $Llist$}
        \State $j, \varphi \gets entry[0], entry[2]$
        \State $\varphi \gets \varphi + (\Delta \phi[j] + \pi)$
        \State $entry[2] \gets \varphi$
    \EndFor

    \State Combine the two lists:
    \State $Glist \gets \text{concatenate}(Llist, \text{reverse}(Rlist))$ \Comment{List with all parameters}

    \State Initialize matrix $A$ of size $M \times M$: 
    \State $A[i,j] \gets i$ for all elements in the $i$th column

    \State Initialize $k \gets 0$ \Comment{Index for $Glist$}
    
    \State Assign entries to $Layers$ (part 1):
    \For{$i \gets 0$ to $M$ with step 2}
        \For{$j \gets 0$ to i-1}
            \State Append $Glist[k]$ to $Layers[A[M-i+j, j]]$
            \State $k \gets k + 1$
        \EndFor
    \EndFor

    \State Assign entries to $Layers$ (part 2):
    \For{$i \gets 1 + (M \% 2)$ to $M$ with step 2}
        \For{$j \gets 0$ to $M - i - 1$}
            \State Append $Glist[k]$ to $Layers[A[j, j+i]]$
            \State $k \gets k + 1$
        \EndFor
    \EndFor

    \State Finalize $Layers$:
    \ForAll{$layer$ in $Layers$}
        \State Sort $layer$ by $layer[0]$ \Comment{Sort by the first element}
        \State Remove the first element of $layer$ \Comment{Remove redundant index}
    \EndFor

    \State \Return $Layers$
\EndFunction
\end{algorithmic}
\end{algorithm}

\section{Acknowledgements}
This work is supported the BMBF project Pho-Quant (grant no 13N16110).

\bibliographystyle{ieeetr}			
\bibliography{biblio}

\end{document}